\documentclass[preprint,english,showpacs,11pt,floatfix]{revtex4}

\usepackage{babel,amsmath,amssymb,dcolumn}
\usepackage[dvips]{graphics}
\usepackage{hyperref}

\usepackage{amsfonts}
\usepackage{amssymb}
\usepackage[dvips]{graphicx}
\usepackage{latexsym}
\usepackage[nooneline]{caption}
\usepackage{subfigure}

\usepackage{dcolumn}
\newcommand{\eear}{\end{eqnarray}}
\newcommand{\be}{\begin{equation}}
\newcommand{\ee}{\end{equation}}
\newcommand{\bc}{\begin{center}}
\newcommand{\ec}{\end{center}}
\newcommand{\bear}{\begin{eqnarray}}

\begin{document}

\title{Interacting Dark Energy and Dark Matter:
observational Constraints from Cosmological Parameters}

\author{Bin Wang, Jiadong Zang}
\email{wangb@fudan.edu.cn} \affiliation{Department of Physics,
Fudan University, 200433 Shanghai}

\author{Chi-Yong Lin}
\email{lcyong@mail.ndhu.edu.tw} \affiliation{Department of
Physics, National Dong Hwa University, Shoufeng, 974 Hualien}

\author{Elcio Abdalla, S. Micheletti}
\email{eabdalla@fma.if.usp.br} \affiliation{Instituto de Fisica,
Universidade de Sao Paulo, C.P.66.318, CEP 05315-970, Sao Paulo}

\begin{abstract}
Several observational constraints are imposed on the interacting
holographic model of Dark Energy and Dark Matter. First we use the
age parameter today, as given by the WMAP results. Subsequently,
we explained the reason why it is possible, as recently observed,
for an old quasar to be observed in early stages of the universe.
We discuss this question in terms of the evolution of the age
parameter as well as in terms of the structure formation. Finally,
we give a detailed discussion of the constraints implied by the
observed CMB low $\ell$ suppression. As a result, the interacting
holographic model has been proved to be robust and with reasonable
bounds predicts a non vanishing interaction of Dark Energy and
Dark Matter.

\end{abstract}


\maketitle

\section{Introduction}

The host of results brought by the continuing operation of WMAP
assures us about the validity of the basis of the Standard
Cosmological Model \cite{wmap1,wmap,wmapcosmos}. While in the
eighties optical observations led us to information concerning the
cosmos with high uncertainties, the present status is based on
observations which deserve the denomination of precision
cosmology.

One of the most tantalizing results connected with the WMAP
observations as well as with the new available Supernova data
\cite{supernova} is the fact that $97\%$ of the Universe consists
of an unknown state of matter, from which $2/3$ is the so called
Dark Energy (DE), responsable for an unexpected cosmological
acceleration and roughly $1/3$ is a Dark Matter (DM), a
gravitationally interacting form of non baryonic matter.

Such forms of energy constitute a major puzzle of modern
cosmology. They attracted a lot of attention and much effort has
been spent to understand them in the last few years. Until now,
the nature and origin of Dark Matter and Dark Energy are still the
source of much debate \cite{padmanabhan}.

Despite the theoretical difficulties in understanding Dark Energy,
independent observational evidence for its existence is
impressively robust. We have three largely independent types of
observational arguments for dark energy: the supernova Hubble
diagram \cite{supernova}, the dynamical evidence for low matter
density, namely the fact that according to inflation there is a
missing component in the energy balance of the universe
\cite{wmapcosmos} and the age of the universe \cite{age}. In
addition, a great success has been scored in high precision
measurements of CMB anisotropy, as well as in galaxy clustering,
the Ly$\alpha$ forest and gravitational lensing
\cite{gravlensing}. Along with these observations, the age of the
universe is one of the most pressing pieces of data disclosing
information about dark energy. Dark energy influences the
evolution of the universe, thus any limit on the age of the
universe during its evolution with redshift will reveal its
nature. As estimated from globular clusters
\cite{globularclusters} and CMB measurements \cite{wmapcosmos},
the total expanding age at $z=0$ is $t_{0}\sim13Gyr$. Such an
astrophysical constraint on the age of the universe at $z=0$ is
useful to limit the equation of state of dark energy \cite{05}.
However, different dark energy models may lead to the same age of
an expanding universe at $z=0$. This degeneracy can be lifted by
examining the age of the universe at different stages of its
evolution and comparing with age estimates of high-redshift
objects. Such a procedure constrains the age at different stages,
being a powerful tool to test the viability of different models
\cite{06}.

On the other hand, Dark Matter is also well established, not only
by the long standing observations of rotation curves in galaxies
\cite{rotationcurves} but also as a result of CMB observations by
WMAP \cite{wmapcosmos}. Although the interpretation of Dark Energy
and Dark Matter as completely independent objects is possible, as
signalized by the CMB results alone which are compatible with the
$\Lambda$CDM model, it is a rather strange and unnatural approach
to the question \cite{pavon,amendola}. Indeed, observationally,
the  $\Lambda$CDM model fits the CMB data alone \cite{wmapcosmos}.
However, the only sensible way such a new state of matter can be
understood at present is by means of a standard introduction of
new fields in the framework of a Quantum Field Theory in a curved
space-time. Even the introduction of a cosmological constant to
explain the acceleration suffers from severe problems to explain
its actual size \cite{cosmoproblem}. Moreover, it does not explain
the fact that today Dark Energy and Dark Matter density fractions
are of the same order of magnitude, the so called coincidence
problem \cite{pavon}. Thus, one might argue that an entirely
independent behavior of DE and DM is very special. Studies on the
interaction between DE and DM have been carried out
\cite{amendola}. It is worthwhile mentioning that the interacting
holographic model was shown to be consistent with the golden SN
data \cite{wga} and can accommodate the transition of the dark
energy equation of state from $\omega_D>-1$ to $\omega_D<-1$
\cite{wga,wla} as recently revealed from extensive data analysis
\cite{11,12}.

The motivation of the present paper is to analyse in detail up to
what point an interaction of the two sectors, namely Dark Energy
and Dark Matter, in the background of the holographic model, is
compatible with observations. We thus begin by briefly explaining
the interacting holographic model and some of its consequences.
Subsequently, we compute the age parameter today and in the past,
giving constraints to the interaction coefficient in order to
explain the existence of an old astrophysical structure, the
quasar APM 0879+5255. Next, as an important complementary piece of
information, we show that such a structure can be generated more
naturally in the interacting model. We further discuss in detail,
by means of a fine numerical analysis and use of the cmbfast code,
the probability that the model explains the low $\ell$ CMB data as
a function of the phenomenological parameters and arrive at the
interesting conclusion that, together with the previous analysis
and with a confidence estimated to be of the order of 90\%, that
the dark energy and the dark matter do interact. In the end we
draw conclusions and try to foresee further developments along
this line.

\section{The Holographic Dark Energy Model}

Standard Quantum Field Theory in a curved background has been
succesfully used in the description of the cosmos. On the other
hand, quantum gravity has never been thoroughly and succesfully
included in a description of a unified theory of all interactions.
Nevertheless, some ideas, such as holography \cite{thooft},
derived from a semi classical description of gravity together with
further general assumptions, have found applications in the
cosmological setup \cite{holo}.

Concerning cosmology, in particular, the energy content of the
universe, we can set up a relation inspired by the fact that the
whole energy content of the cosmos cannot exceed the mass of a
Black Hole with the same size of the universe, which we call $L$.
We thus surpose that \be \rho_D = \frac {3c^2M_p}{L^2} \ee where a
phenomenological constant $c$ has been conveniently introduced,
characterizing a free parameter of our model while $M_p$ is the
Planck mass.

Such a formulation is known in the literature as the holographic
hypothesis \cite{holohypothesis}. In order not to violate the
second law of thermodynamics in the event horizon, it has been
argued that $c\ge \sqrt {\Omega_\Lambda}$, or in general $c\ge 1$
\cite{miaoli}. Since the thermodynamics in the event horizon may
be problematic \cite{wanggongabdth} and the IR regulator might not
be simply related to the future event horizon
\cite{holohypothesis} \cite{shen} \cite{huang}, we suppose that
there is a lower bound for $c$ but try not to specify it strictly.
We shall see however that  a lower bound is natural (see also
\cite{wga}). In most of the paper we suppose that $c=1$. As we
shall see later, bounds in $c$ are not very restrictive.

As far as energy conservation is concerned, we suppose that the
interaction is described by the (separately non conserving)
equations \bear \dot\rho_m + 3H\rho_m =+Q\nonumber\\ \dot\rho_D +
3H(1+\omega_D)\rho_D =-Q\label{interaction} \eear where $Q$ is
some interaction. For the moment we take for granted that the
interaction is the one proposed on general grounds in
\cite{pavonandandzim}, that is, \be Q=3b^2H(\rho_m+\rho_D) \ee
where $b^2$ is a second phenomenological constant coupling DE and
DM. It has to be fit by observational data. The model has been
further discussed in \cite{wga} for a flat universe and \cite{wla}
for the closed universe. The question has been further discussed
in \cite{shen,huang}.

We just consider the interaction between DE and DM, but neglect
the interaction between DE and Baryonic matter. The minimal
coupling between DE and Baryonic matter is usually assumed in
order to avoid testable violations of the equivalence principle
\cite{Ame}  and a possible conflict with observational constraints
on long-range forces \cite{Ser} . In \cite{pavonxx}, it was found
that good consistency with the observational bounds for a variety
of parameter combinations does not depend on whether or not the
baryons are included in the interaction. We thus assume that
baryonic matter is minimally coupled in our study.

Finally, we should mention that in general we admit that the
energy pressure relation $\omega_D =p_D/\rho_D$ can be redshift
(time) dependent, which is natural from the point of view of the
two fluids interacting, as described in (\ref{interaction}). This
is also consistent with the recent data analysis \cite{11,12}.

\section{Age Constraints}

In this section we consider the age of the universe, as influenced
by the holographic model interaction.

\begin{figure}[t] \label{fig1}
\begin{center}
\includegraphics[width=10cm]{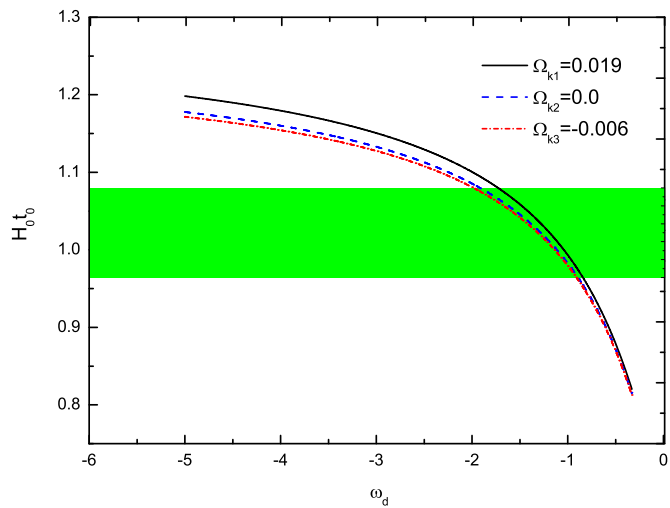}
\end{center}
\caption{ The age of the universe as a function of the (constant)
parameter defined by the equation of state. The shadowed region is
the total age at $z=0$ got by WMAP\cite{wmapcosmos}} \end{figure}

To begin with, we consider the simple cosmic expansion described
by the Friedmann equation, \begin{equation}\label{eq1}
     H^2=H^2_0[\Omega_{m_0}(1+z)^3+
     \Omega_{D_0}f(z)-\Omega_{k_0}(1+z)^2]\quad , \end{equation} where $\Omega_{m_0}$,$\Omega_{D_0}$ and $\Omega_{k_0}$ refer to densities of matter, dark energy and curvature at the present day in units of the critical density. The function  $f(z)$ is related to the equation of  state of dark energy by $f(z)=e^{3\int_0^z \frac{1+\omega(z')}{1+z'}dz}$. The age of the universe at redshift $z$ is \begin{equation}\label{eq2}
     t(z)=\int_z^\infty \frac{dz'}{(1+z')H(z')}\quad .
\end{equation}

WMAP three year results \cite{wmapcosmos} tell us that at $z=0$,
$t_0=13.73^{+0.13}_{-0.17}Gyr$ and that the current value of the
Hubble parameter is $H_0=73.4^{+2.8}_{-3.8}km/s/Mpc$, which are
compatible with direct age estimates from globular clusters
\cite{globularclusters} and HST measurements \cite{13} for the
Hubble parameter. Putting in $H_0$-independent terms one finds
$0.9646\le H_0t_0\le 1.0794$. Using this WMAP range of the
dimensionless age parameter at present, we can put constraints on
the dark energy. By considering the constant equation of state in
Fig. 1, we have shown the age constraints on the dark energy
equation of state for flat space, closed and open spaces with
curvatures allowed by WMAP observations \cite{wmapcosmos}. For the
flat universe, $\Omega_{k_2}=0$, we obtain the allowed range
$\omega_D\in[-1.9099,-0.8939]$; for a closed universe with
$\Omega_{k_1}=0.019$ we get $\omega_D\in[-1.7221,-0.8427]$;
finally, for a flat universe with $\Omega_{k_3}=-0.006$,
$\omega_D\in[-1.9777,-0.9111]$. Current universe age constraints
are all consistent with the combination of WMAP, large scale
structure and supernova data, that is,
$\omega_D=-1.06^{+0.13}_{-0.08}$ \cite{wmapcosmos}.

We analyse now  the old quasar APM 0879+5255, inquiring about the
constraints such data can bring to the question of Dark Energy.
This quasar was discovered at a redshift $z=3.91$ and its age has
been estimated to be $2.1Gyr$ \cite{08,06}. Employing the WMAP
determination of the Hubble parameter, the age of the quasar is in
the dimensionless interval $0.148\le Tg\le 0.162$. A viable
cosmological model should predict a considerably older universe at
that high redshift in order to be compatible with the existence of
this object.

\begin{figure}[t]
   \centering
     \subfigure[\hspace{0.1cm}]
     {\includegraphics[width=0.4\textwidth]{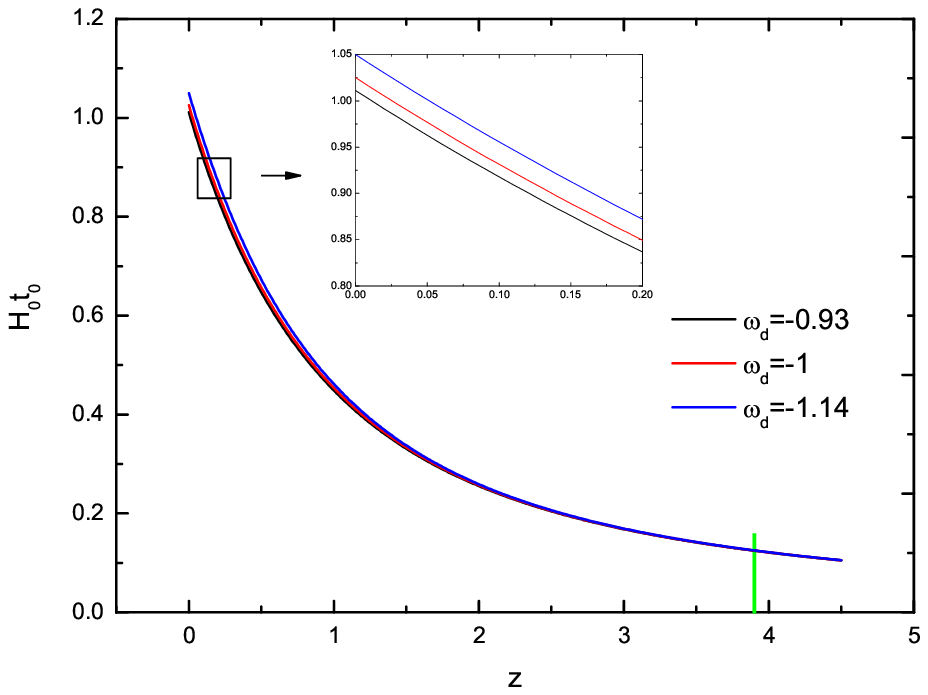}\label{fig2a}}
     \hspace{0.01\textwidth}
     \subfigure[\hspace{0.1cm}]
     {\includegraphics[width=0.4\textwidth]{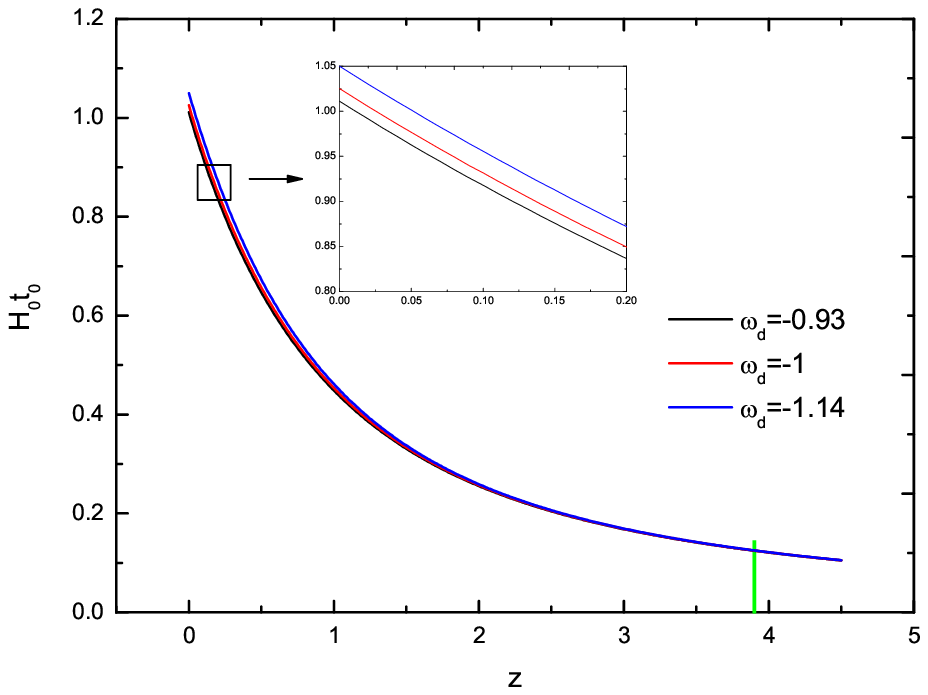}\label{fig2b}}\\
     \subfigure[\hspace{0.1cm}]
     {\includegraphics[width=0.4\textwidth]{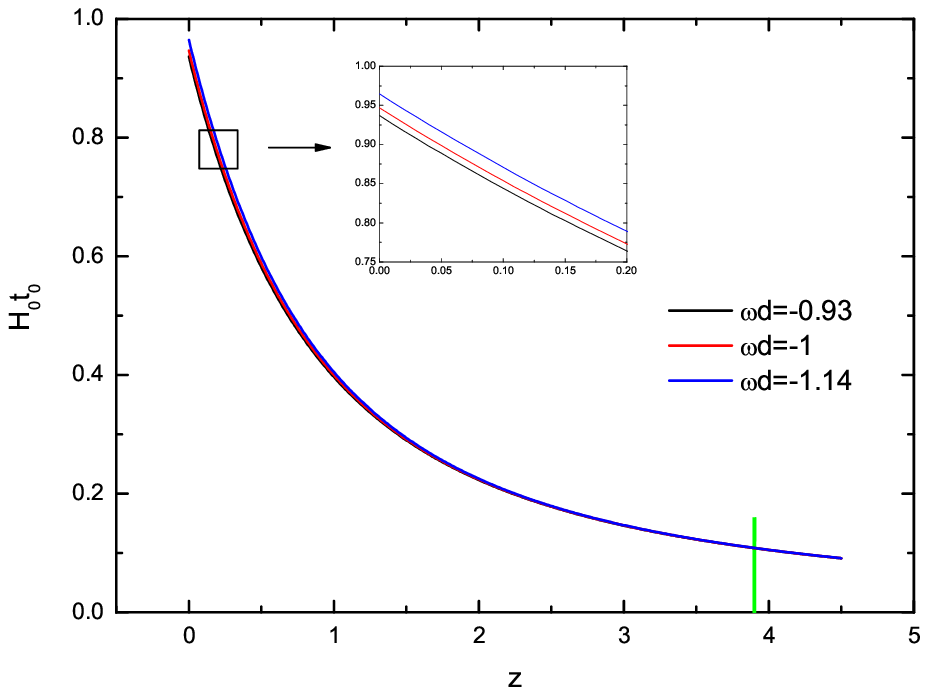}\label{fig2c}}
     \hspace{0.01\textwidth}
     \subfigure[\hspace{0.1cm}]
     {\includegraphics[width=0.4\textwidth]{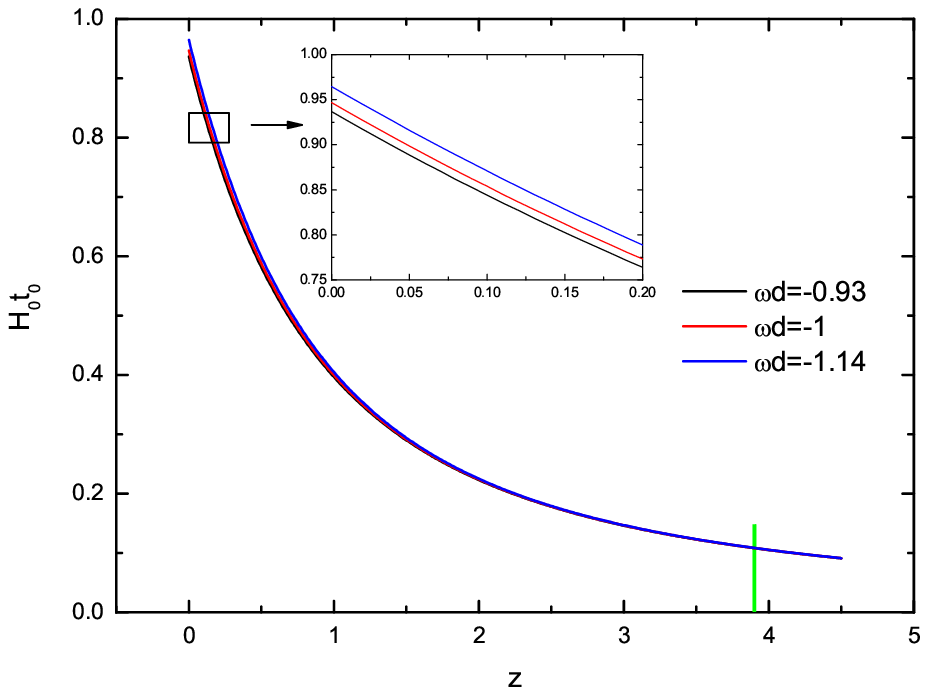}\label{fig2d}}
     \caption{Dimensionless age parameter as a function of redshift
     for simple dark energy models. We have takenthe following
     parameters:
$H_0=76.2 Km/s/Mpc, \Omega_D=0.76$ in (a), $H_0=69.6 Km/s/Mpc, \Omega_D=0.76$ in (b), $H_0=76.2 Km/s/Mpc, \Omega_D=0.68$ in (c) and $H_0=69.6 Km/s/Mpc, \Omega_D=0.68$ in (d). All curves cross the shadowed area yielding an age parameter smaller than the value $2.1 Gyr$ required by the quasar APM 08279+5255.  } \label{fig2} \end{figure} 

Figs. \ref{fig2a}-\ref{fig2d} show the dimensionless age parameter
of a flat universe as a function of the redshift with different
values of equation of state of dark energy allowed by the
observations. In order to assure the robustness of our result, we
have adopted in our computation the upper and lower limits of
$H_0$ and $\Omega_{D_0}$ according to the WMAP results
\cite{wmapcosmos}. We found that all curves cross the shadowed bar
at $z=3.91$, thus yielding an age parameter smaller than the value
$2.1Gyr$ required by the quasar APM 0879+5255. This result also
holds for a universe with a small positive curvature, permitted by
WMAP observations. Therefore, the simple cosmological model with
dark energy is not compatible with the age estimate of the old
quasar. The same result was also found for other dark energy
models by using the HST constraint on the Hubble parameter
\cite{06}.

\begin{figure}[t]
   \centering
     \subfigure[\hspace{0.1cm}]
     {\includegraphics[width=0.4\textwidth]{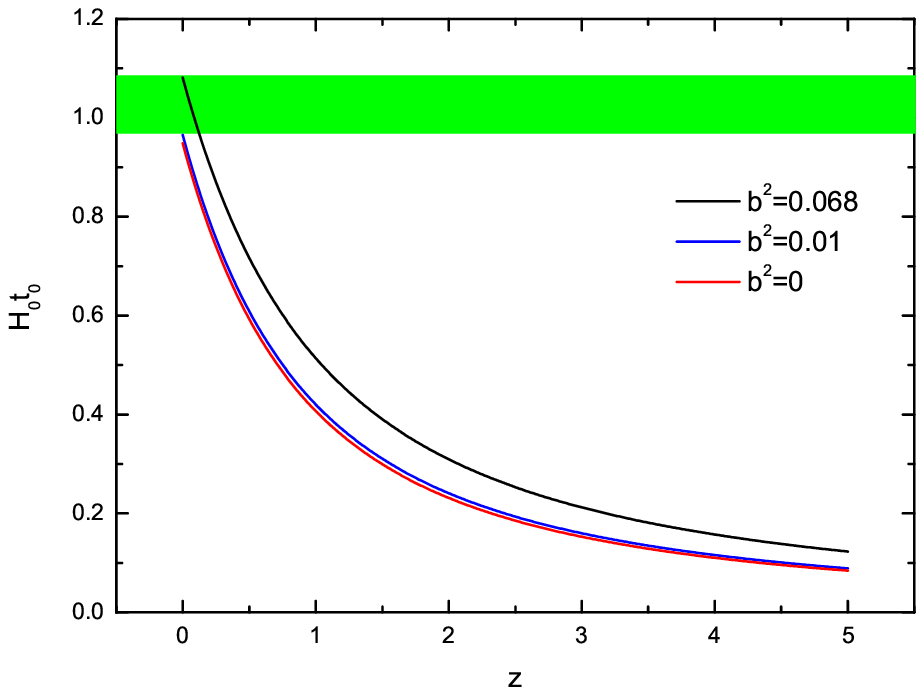}\label{fig3a}}
     \hspace{0.01\textwidth}
     \subfigure[\hspace{0.1cm}]
     {\includegraphics[width=0.4\textwidth]{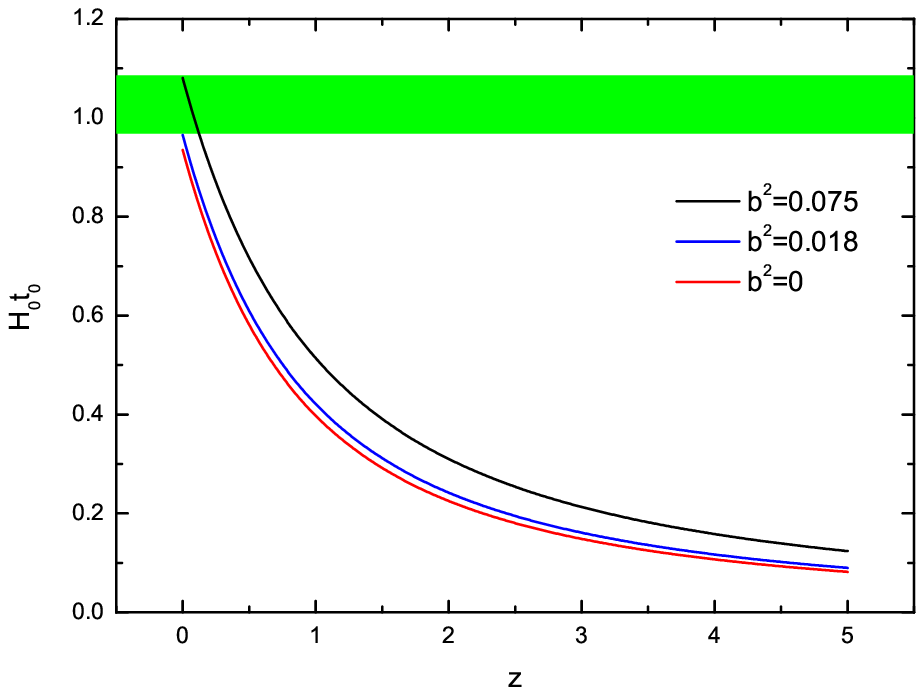}\label{fig3b}}
     \caption{Age of the universe as a function of the redshift
     for interacting holographic dark energy models. The shadowed
     region is the total age at $z=0$ observed from WMAP
     \cite{wmapcosmos}. In (a)
     we considered $\Omega_k=0$; in (b) $\Omega_k=0.019$.} \label{fig3} \end{figure} 
We consider now the interacting holographic dark energy model
proposed in \cite{wga,wla}. With the interaction, neither dark
energy nor dark matter can conserve and evolve separately. For the
closed universe, the evolution behavior of the dark energy was
obtained as \cite{wla}, \begin{equation}\label{eq3}
     \frac{\Omega'_D}{\Omega^2_D}
\frac{(1-\Omega_k-\Omega_D)}{(1+\Omega_k)}
     [\frac{2cosy}{c\sqrt{\Omega_D}}
     +\frac{1}{\Omega_D}+\frac{\Omega'_k}{\Omega_D
     (1+\Omega_k-\Omega_D)}-\frac{3b^2(1+\Omega_k)}
     {\Omega_D(1+\Omega_k-\Omega_D)}]\quad , \end{equation} where $b^2$ and $c$ have been defined in section 2. The prime denotes the derivative with respect to $x=lna$ and $cosy=\sqrt{1-\frac{c^2 \Omega_k}{\Omega_D}}$. The equation of state of dark energy was expressed as \cite{wla} \begin{equation}\label{eq4}
     \omega_D=-\frac{1}{3}-
     \frac{2\sqrt{\Omega_D}}{3c}cosy+
     \frac{b^2(1+\Omega_k)}{\Omega_D}\quad .
\end{equation}
The evolution of the Hubble parameter was derived as well,
\begin{equation}\label{eq5}
     \frac{H'}{H}=-\frac{3\Omega_D(1+\omega_D+r)}{2}+\Omega_k,
\end{equation}
where $r=\frac{1+\Omega_k-\Omega_D}{\Omega_D}$ is the ratio of the
energy densities. Appropriately choosing the coupling between dark
energy and dark matter, this model can also accommodate the
transition of the dark energy equation of state from $\omega_D>-1$
to $\omega_D<-1$ \cite{wla,wga}, which is in agreement with the
recent analysis of the type Ia supernova data \cite{11,12}.

We now want to limit this interacting holographic dark energy
model from age considerations. Employing the age of the expanding
universe at $z=0$, we have shown our results in Fig.\ref{fig3a}
and Fig.\ref{fig3b} respectively for $\Omega_k=0$ and
$\Omega_k=0.019$, the closed universe allowed by WMAP. For
$\Omega_k=0$, the limit on the age of the universe from WMAP
\cite{wmapcosmos} puts the constraint on the coupling between dark
energy and dark matter in the interval $0.01\le b^2\le 0.068$ for
$c=1$. For $\Omega_k=0.019$, the allowed coupling between dark
energy and dark matter can be got at $0.018\le b^2\le 0.075$ for
$c=1$ from the present age of the universe.

\begin{figure}[t]
   \centering
     \subfigure[\hspace{0.1cm}]
     {\includegraphics[width=0.4\textwidth]{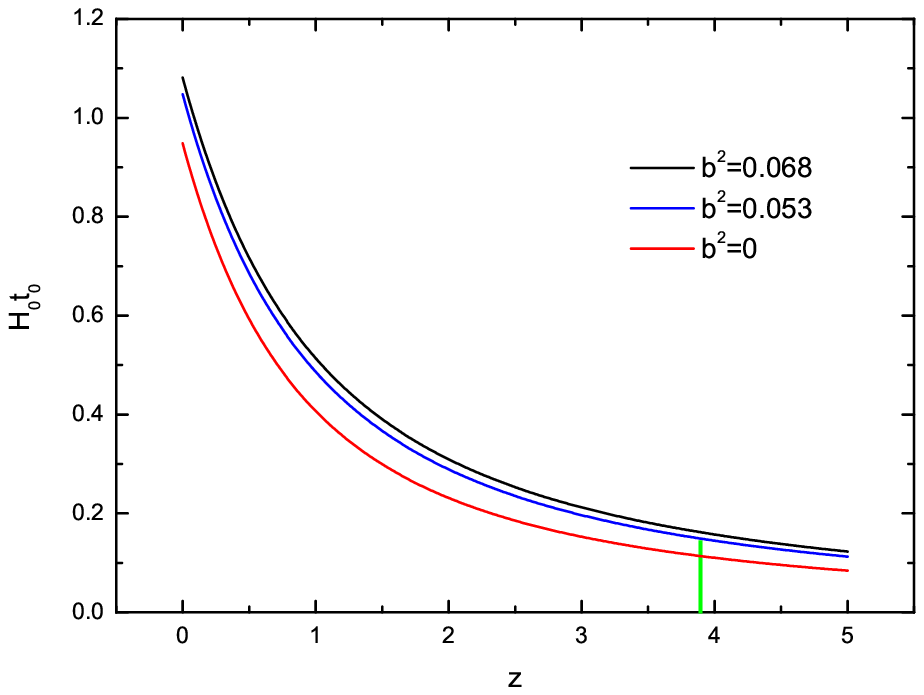}\label{fig4a}}
     \hspace{0.01\textwidth}
     \subfigure[\hspace{0.1cm}]
     {\includegraphics[width=0.4\textwidth]{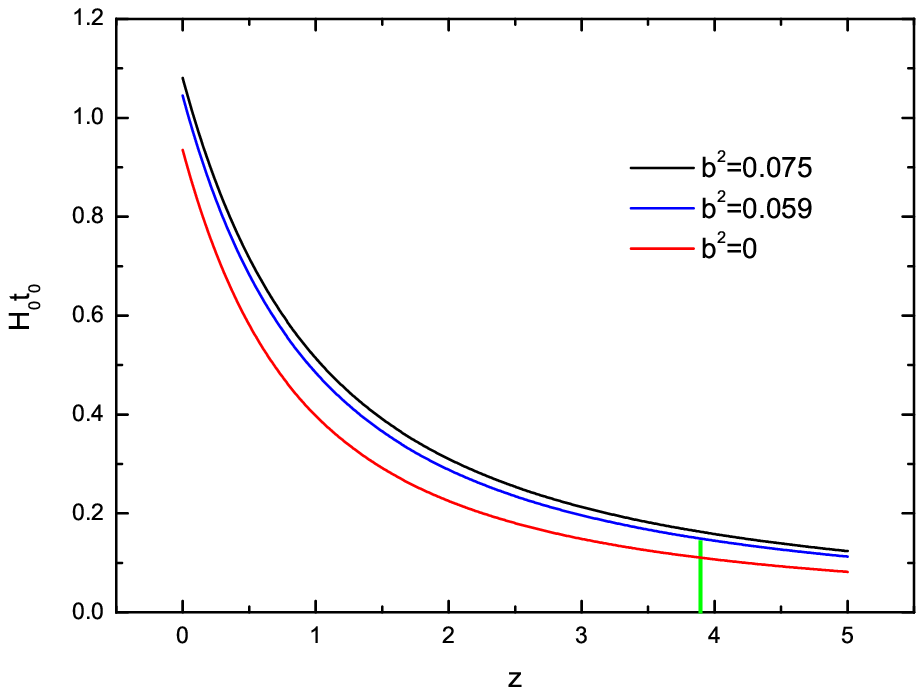}\label{fig4b}}
     \caption{Dimensionless age parameter as a function of
     redshift for interacting holographic dark energy models.
     In (a) we considered $\Omega_k=0$, in (b) $\Omega_k=0.019$.
     We see that for appropriately coupling between dark energy
     and dark matter, the interacting holographic dark energy
     model can accommodate the existance of APM08279+5255 system.}
\label{fig4}
\end{figure}
The estimated age of an old quasar APM 0879+5255 at redshift
$z=3.91$ has also been used to test the viability of the
interacting holographic dark energy model. By adopting the lower
and upper bounds of values of Hubble parameter and dark energy
density respectively from WMAP \cite{wmapcosmos}, $H_0=69.6$,
$\Omega_D=0.72$, we have shown in Fig.\ref{fig4a} and
Fig.\ref{fig4b} that the interacting holographic dark energy model
with appropriate coupling between dark energy and dark matter is
compatible with the estimated age for the APM 0879+5255. For
$\Omega_k=0$, the appropriate coupling is required as $b^2>0.053$
for $c=1$ and for $\Omega_k=0.019$, $b^2>0.059$ for $c=1$ in order
to accommodate the existence of the old quasar. Thus we have shown
that the interacting holographic dark energy model is viable from
the age constraints.

\begin{figure}[t]
   \centering
     \subfigure[\hspace{0.1cm}]
     {\includegraphics[width=0.4\textwidth]{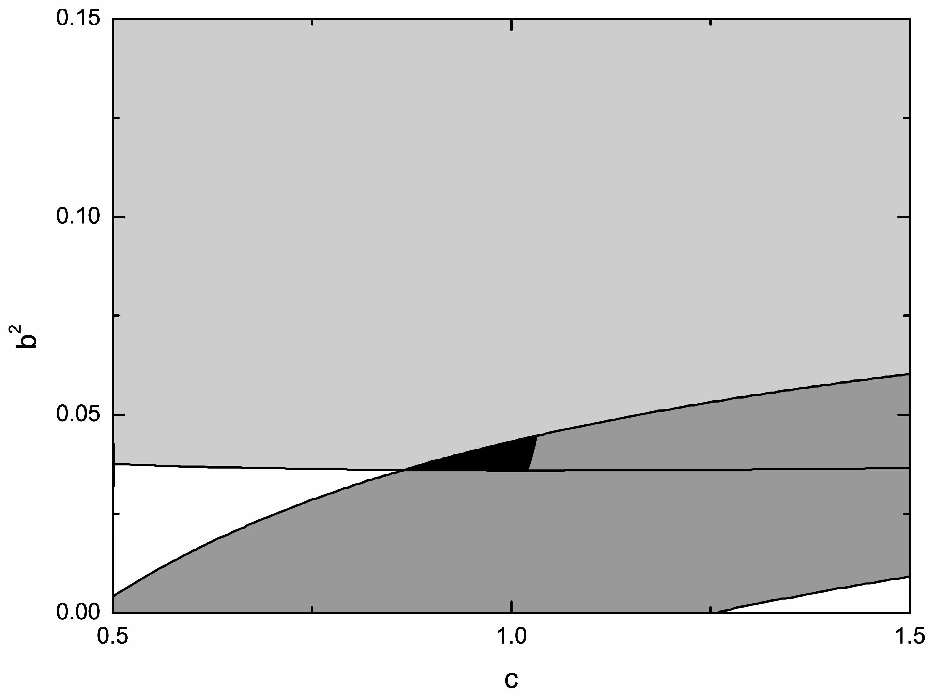}\label{fig5a}}
     \hspace{0.01\textwidth}
     \subfigure[\hspace{0.1cm}]
     {\includegraphics[width=0.4\textwidth]{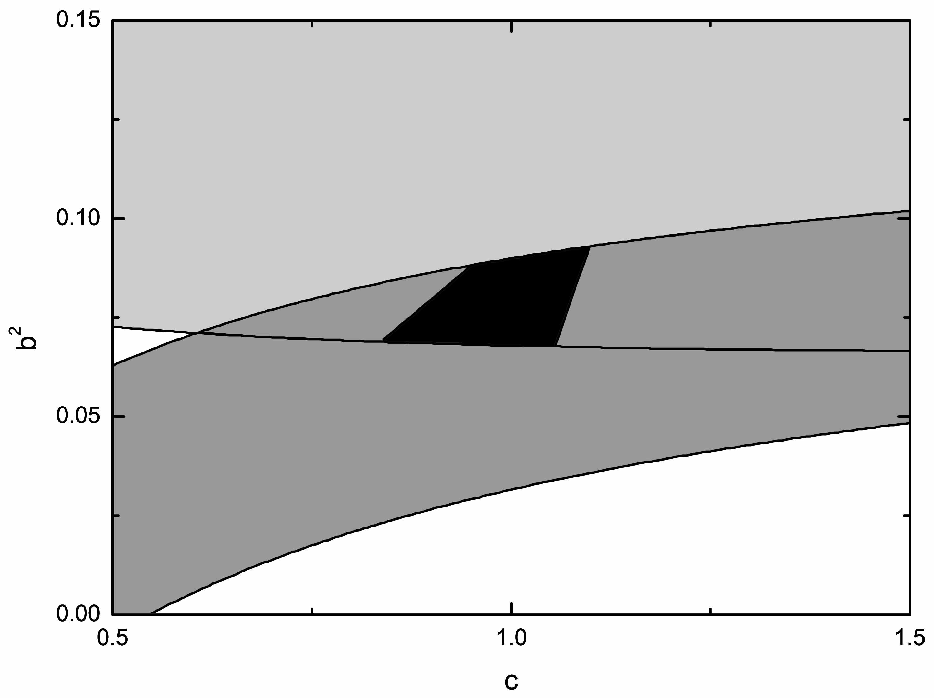}\label{fig5b}}\\
     \subfigure[\hspace{0.1cm}]
     {\includegraphics[width=0.4\textwidth]{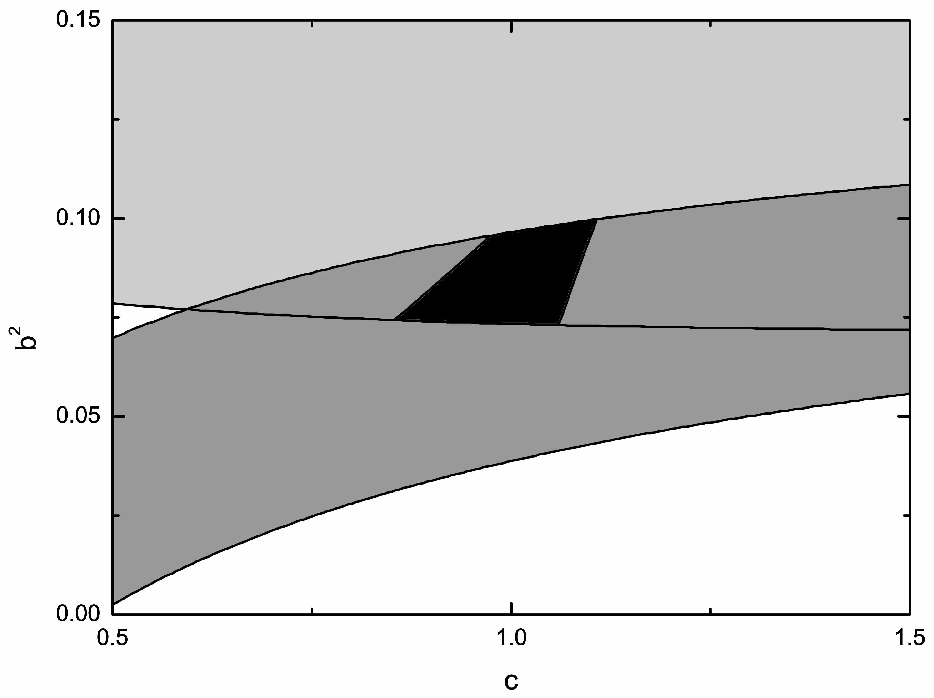}\label{fig5c}}
     \hspace{0.01\textwidth}
     \subfigure[\hspace{0.1cm}]
     {\includegraphics[width=0.4\textwidth]{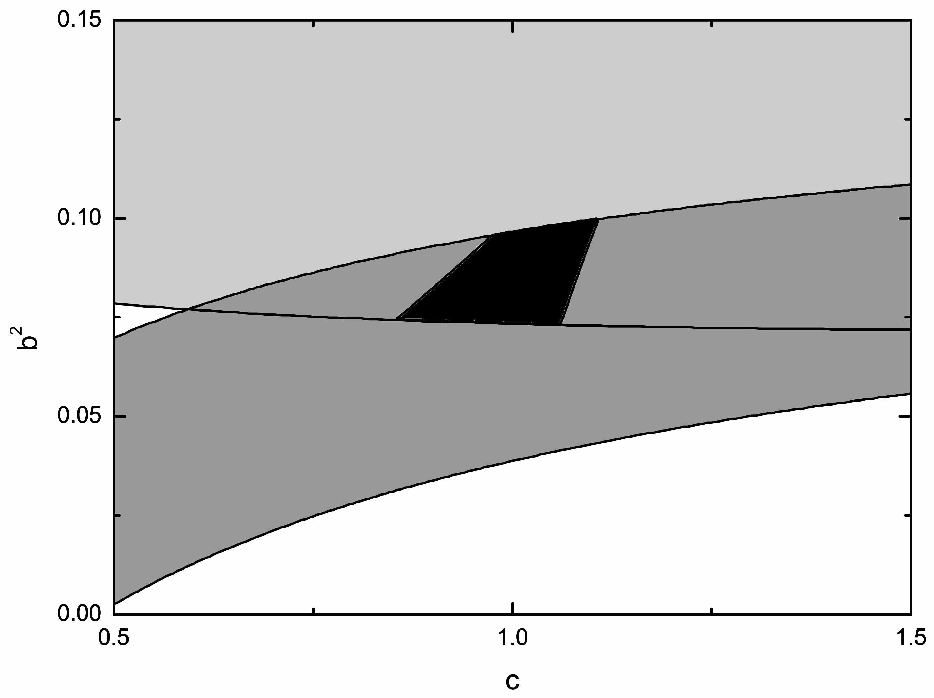}\label{fig5d}}
     \caption{The constrained parameter space of $b^2$ and $c$.
     The dark grey is the constraint from the total age at
     $z=0$ from WMAP \cite{wmapcosmos},
     the light grey is the constraint from the old quasar
     APM08279+5255 \cite{06,08}
     and the dark region is the parameter space compatible with
     the $w_D$ crossing $-1$ \cite{12} and the age constraints.
     We have taken $\Omega_k=0, \Omega_D=0.76$ in (a), $\Omega_k=0,
     \Omega_D=0.68$ in (b),
     $\Omega_k=0.019, \Omega_D=0.76$ in (c) and $\Omega_k=0.019,
     \Omega_D=0.68$ in (d).}
\label{fig5}
\end{figure}
Combining with the requirement that $\omega_D$ crosses -1 and the
age constraints from $z=0$ and $z=3.91$, we have obtained the
allowed parameter space of $b^2$ and $c$ in Fig.\ref{fig5}. Within
the black region of parameter space, the interacting holographic
dark energy model is compatible with the transition behavior of
$\omega_D$ and the age constraints from observation. It can
describe the accelerated expansion of our universe happened before
the present era.

\section{Structure Formation}

In a model with interaction we certainly expect that structure
formation has a different fate as compared with the non
interacting case. In the model defined by (\ref{interaction}) dark
matter is continuously being fed at the expense of dark energy and
the clumping properties have obviously to change.

In such a framework we shall discuss the formation of a structure,
such as the quasar previously discussed, in the young universe. We
thus have to consider the matter density perturbation \be \delta
\equiv \frac{\delta\rho_m}{\rho_m} \ee and its time evolution. We
do not consider dark energy perturbations, assuming that it does
not clump sufficiently to help forming structures. Rather we
assume that hadronic matter just follows the general dark matter
pattern. Thus, within general relativity the evolution equation of
dark matter inhomogeneities is given by
\begin{equation}\label{eq6}
     \ddot\delta+2H\dot\delta-\frac{a\delta}{2}
     \frac{d(H^2\Omega_m)}{da}=0\quad .
\end{equation}
If there is no interaction, $b^2=0$ and we know that
$\Omega_m=\frac{\rho_m}{3H^2}=\frac{\Omega_{m_0}H^2_0}{H^2a^3}$.
In that case the equation above becomes the usual matter density
perturbation equation in a simple cosmological set up \cite{14},
\begin{equation}\label{eq7}
     \ddot\delta+2H\dot\delta-\frac{3}{2}H^2\Omega_m\delta=0\quad .
\end{equation}
Changing variables from $t$ to $x=\ln a$, Eq. (\ref{eq6}) can be
rewritten as
\begin{equation}\label{eq8}
     \frac{d^2\delta}{dx^2}H^2+\frac{d\delta}{dx}[2H^2+
     \frac{dH}{dx}H]
     +\frac{\delta}{2}\frac{d(H^2\Omega_m)}{dx}=0\quad .
\end{equation}
Defining the growth variable $G=\delta/a$, Eq(\ref{eq8}) can be
further rearranged as
\begin{equation}\label{eq9}
     \frac{d^2 G}{d^2 x}+(4+\frac{1}{H}\frac{dH}{dx})\frac{dG}{dx}
     +(3+\frac{1}{H}\frac{dH}{dx}+\frac{1}{2H^2}
     \frac{d(H^2\Omega_m)}{dx})G=0\quad .
\end{equation}
\begin{figure}[t]
   \centering
     \subfigure[\hspace{0.1cm}]
     {\includegraphics[width=0.4\textwidth]{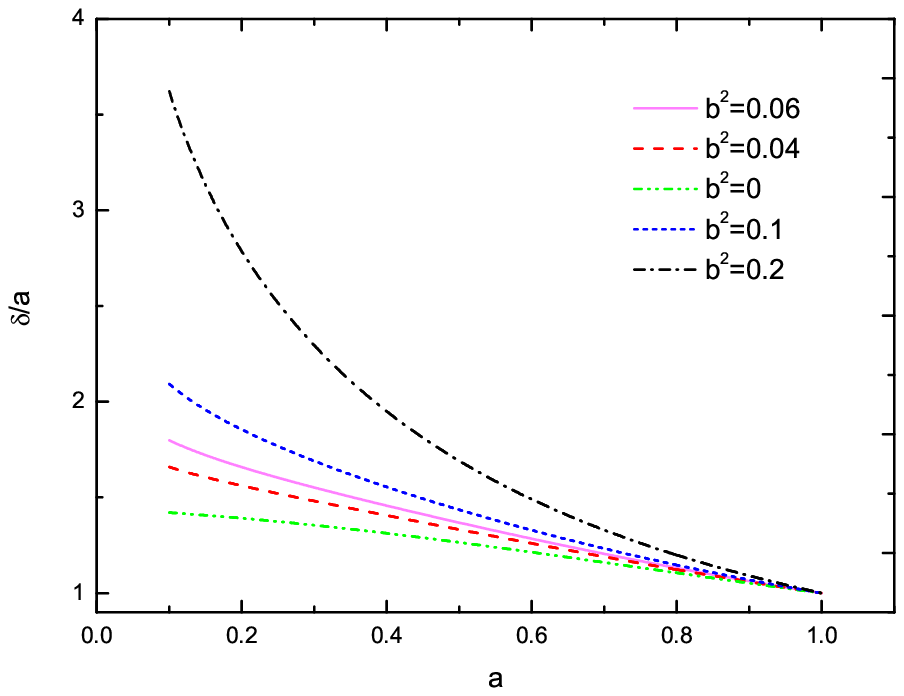}\label{fig6a}}
     \hspace{0.01\textwidth}
     \subfigure[\hspace{0.1cm}]
     {\includegraphics[width=0.4\textwidth]{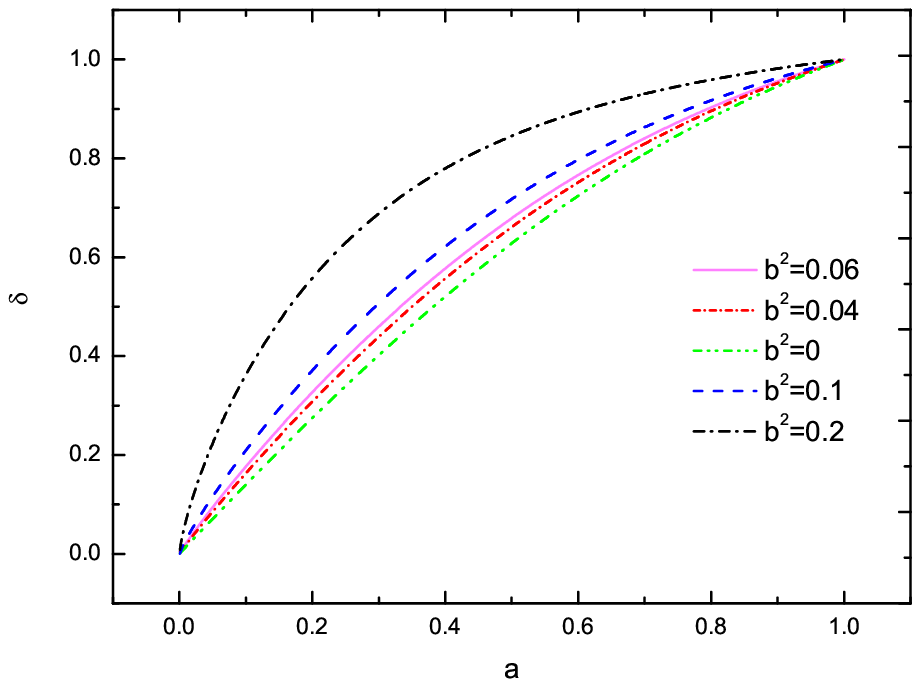}\label{fig6b}}
     \caption{Evolution behavior of the matter density perturbation.}
\label{fig6}
\end{figure}
Normalizing the density perturbation in terms of the present
amplitude, we show the solution to the growth variable in
Fig.\ref{fig6a}. Plotting the matter density perturbation
evolution in Fig.\ref{fig6b}, we found that with strong coupling
between dark energy and dark matter, the matter density
perturbation is stronger during the universe evolution till today,
which shows that the interaction between dark energy and dark
matter enhances the clustering of dark matter perturbation
compared to the noninteracting case in the past. This phenomenon
was also observed by studying quintessence interacting with dark
matter \cite{15}. The strong clustering of dark matter accounts
for the old quasar appearing in the young universe with
interacting dark energy. From Fig.6a we see that for stronger
coupling between dark energy and dark matter, the growth decreases
faster. This is due to the fact that for stronger coupling between
dark energy and dark matter, dark energy will reveal itself
earlier, the universe will evolve earlier into the accelerated
expansion, and $\omega_D$ can even cross $-1$ and stay at
$\omega_D<-1$ earlier \cite{wga,wla}. Thus the stronger repulsive
pressure from dark energy in the early time for bigger $b^2$
implies a faster decrease of $\delta/a$. If we observe the matter
density perturbation in the future, the stronger coupling between
dark energy and dark matter will weaken the clustering of the dark
matter perturbation and it will be more difficult for the
structure to be formed.

\begin{figure}[t]
   \centering
     \subfigure[\hspace{0.1cm}]
     {\includegraphics[width=0.4\textwidth]{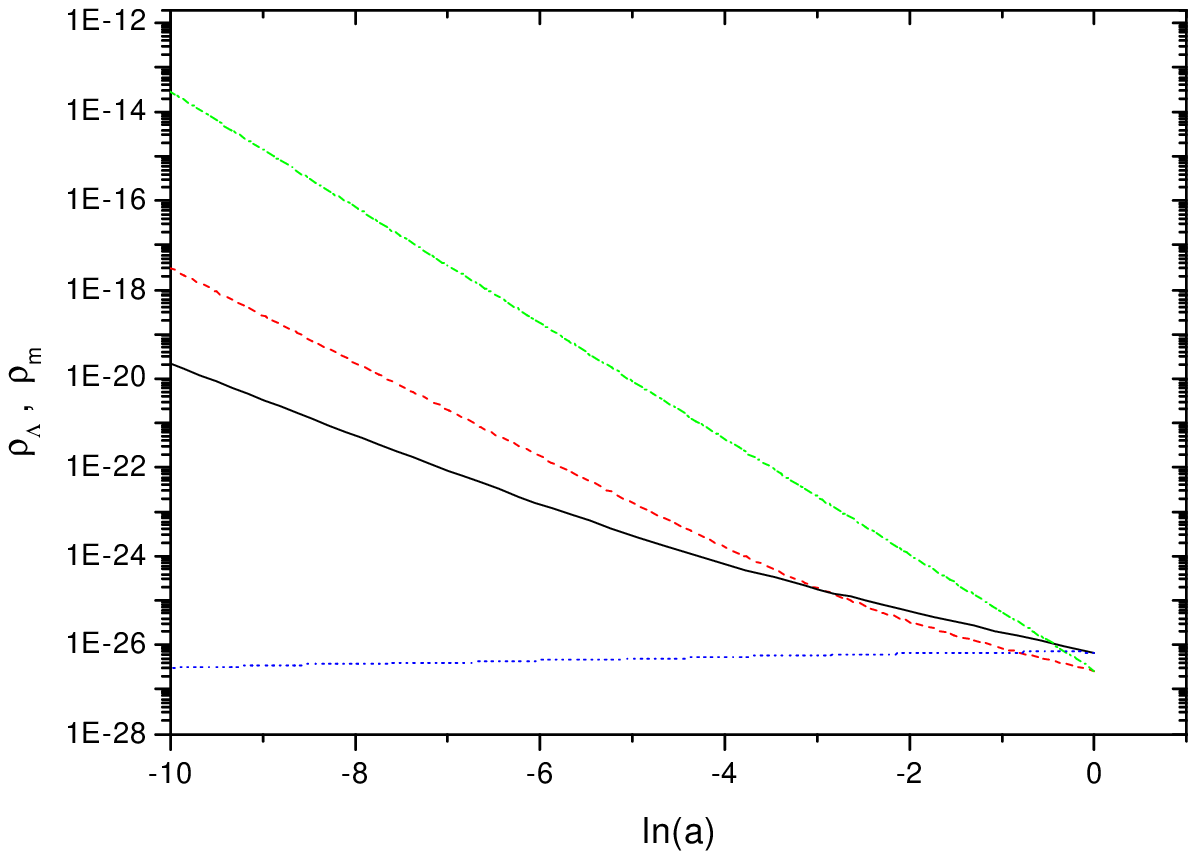}}
     \hspace{0.01\textwidth}
     \subfigure[\hspace{0.1cm}]
     {\includegraphics[width=0.4\textwidth]{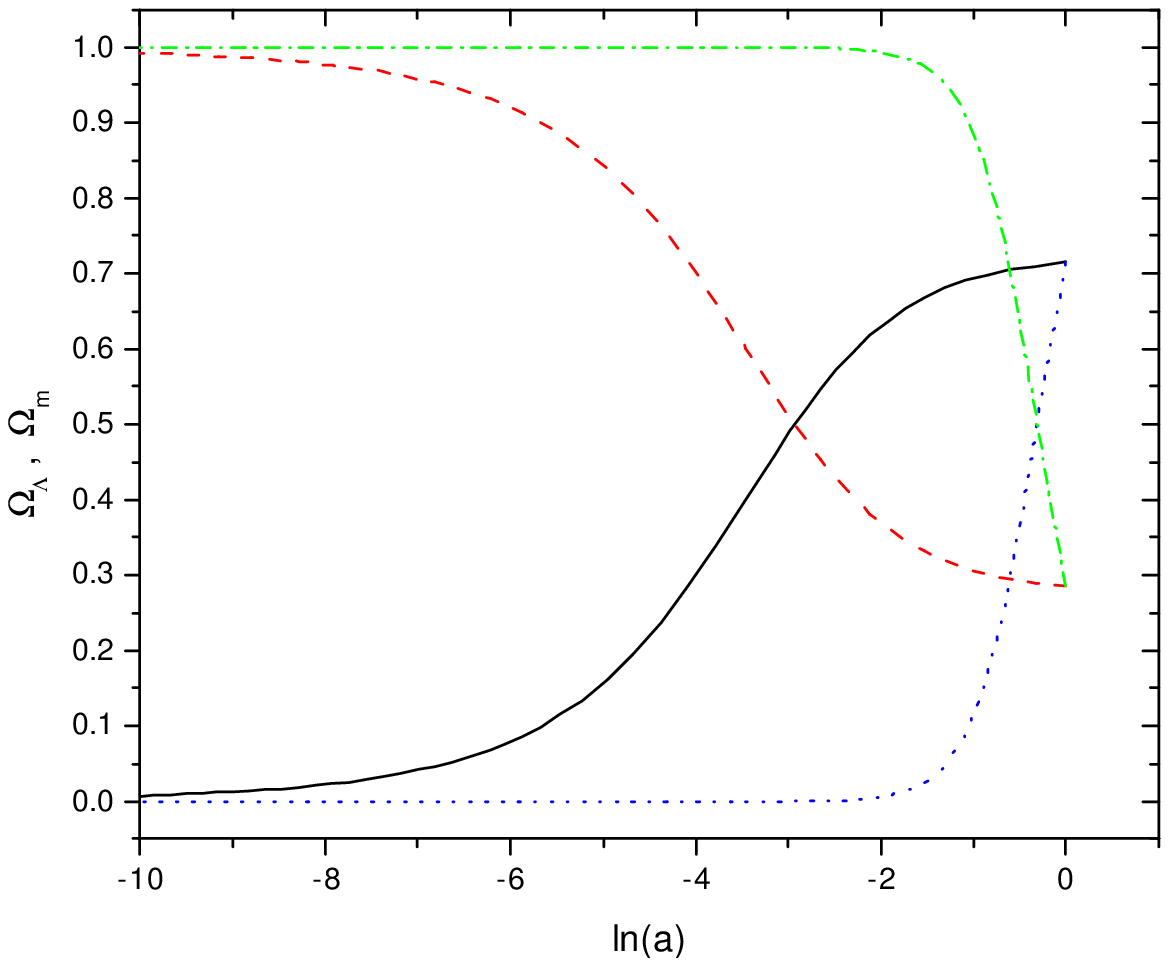}}
     \caption{Evolution of Dark Energy and Dark Matter,
     with or without the interaction term. The line descending
     quickest corresponds to dark matter without interaction.
     The other lines, by order of steepness, are dark matter
     with interaction, dark energy with interaction and dark energy
     without interaction, respectively. The same type of lines
     describe the corresponding evolutions for corresponding
     $\Omega$'s (as a matter of fact, we are using here
     the parametrization I, given below, with $b^2 = 0.18$. The
     qualitative results do not
     depend on this assumption).}
\label{fignew1}
\end{figure}

With interaction Dark Energy and Dark Matter follow one another,
as displayed in figure (\ref{fignew1}). This means that in the
recent history of the universe dark energy is being transformed
into dark matter and the fluctuations do get more effective in the
past according to figure 6. Therefore, in the beginning there were
the same fluctuations, then dark matter was enhanced leading to
higher fluctuations, to finish now with the same structures, but
with more structures in the past, such as the case of the old
quasar. Note that the stronger  the interaction, the more
effectively structures will have been formed in the past. We thus
conclude that we must have $b^2 \gtrsim 0.05$ in this model.

The structure formed as the quasar APM 0879+5255 is thus a further
support of the interaction between these rather puzzling objects
that may open an immense avenue for the study of the universe.

\section{Numerical Analysis of low $\ell$ CMB spectrum}

In our previous analysis for a flat universe \cite{wga} as well as
in a closed universe \cite{wla} we obtained several constraints on
the parameters of the interacting holographic model. In this
section we consider the consequences of the detailed analysis of
the low $\ell$ suppression of the CMB power spectrum revealed by
COBE/WMAP for the phenomenological constants that have been
defined. The question of small $\ell$ suppression was investigated
for flat \cite{shen} as well as closed \cite{huang} universes.

Here we will perform a more detailed numerical analysis in order
to probe how essential is the coupling between DE and DM. We are
going to show that with a $90\%$ confidence level the interaction
is non vanishing, which constitutes a further support of the
statements collected in the previous sections. This opens up a
wide road for models of DE/DM. It also signalizes that a pure
cosmological constant possibly cannot describe such dynamics of
those essential parts of the universe.

As already discussed in \cite{wga,wla,shen,huang} we interprete
the infrared cutoff as the maximum possible wavelenght,
$\lambda_c=2L$. Since $\Omega_{\Lambda}=\rho_{\Lambda}/\rho_c$ and
$\rho_c 3M_p^2H^2$, we have \bear k_c =\frac{\pi}{c}H^0
\sqrt{\Omega_{\Lambda}^0}. \eear The above expression represents
the minimum wave number allowed for the computation of the power
spectrum. Thus we obtain, for the spectral coefficient, \bear
C_\ell = (4\pi^2)\int_{k_c}^\infty k^2dkP_{\Psi}(k)|\Delta
T_\ell(k,\tau=\tau_0)|^2, \eear where $P_{\Psi}(k) $ is the
initial power spectrum and $\Delta T_\ell(k,\tau\tau_0)$ is the
$\ell$th coefficient.

We use the  CMBFAST programm  version 4.5.1, modified to take into
account the cutoff $k_c$.

Since we are lack of the knowledge of the perturbation theory in
including the interaction between DE and DM, in fitting the WMAP
data by using the CMBFAST we will first estimate the value of $c$
without taking into account the coupling between DE and DM. The
interaction becomes important for values of $a$ near $1$ (today).
Considering the equation of state of DE is time-dependent as
disclosed by recent data analysis\cite{11,12}, we will adopt two
extensively discussed DE parametrization models, \bear
\label{eqest1} \omega^I (z) &=& \omega_0 + \omega_1 \frac{z}{z +
1}  \quad , \\
\label{eqest2} \omega^{II} (z) &=& \omega_0 + \omega_1 \frac{z}{(z
+ 1)^2}\quad . \eear

From (\ref{interaction}), we know that the ratio of energy
densities $r = \rho_m/\rho_D $ obeys  (in this section we only
consider the flat case) \be \dot{r} = 3b^2H(1 + r)^2 + 3Hr\omega_D
\quad . \ee Using the Friedmann equation $\Omega_m + \Omega_D = 1$
(valid for a flat universe) as well as $\dot{r}= -\dot{\Omega}_D /
\Omega_D^2 $, we arrive at \be \omega_D = - \frac{\Omega_D
'}{3\Omega_D (1 - \Omega_D ) }-\frac{b^2}{\Omega_D (1 - \Omega_D
)}    \quad . \ee After some algebra, we get \be \label{eqestint}
\omega_D = - \frac{1}{3} - \frac{2\sqrt{\Omega_D}}{3c} -
\frac{b^2}{\Omega_D}\quad  . \ee

Using the modified CMBFAST 4.5.1 (given in www.cmbfast.org)
\cite{seljak}, we can again study the effect of the holographic
model on the small $\ell$ CMB spectrum. We require that the
equation of state of the holographic model persists with the same
behavior as that of the two parametrizations above respectively.

We want to find a set of cosmological parameters maximizing the
likelihood function  $\mathcal{L}$.  However, since the cut
parameter $c$ only affects the spectral region of low multipoles
($l < 10$), such a parameter has negligible correlation with the
other relevant parameters, thus we seek at finding the value of
$c$ maximizing the function $\mathcal{L}(1/c)$.

For parametrization $I$ we used the parameters given in table 1
\cite{seljak}.

\textbf{Table 1}: cosmological parameters for a flat universe with
equation of state $I$

\begin{center}
\begin{tabular}{|c|c|}
\hline $\Omega_{\Lambda }$
& $0.715^{+ 0.023+0.045+0.066}_{- 0.024-0.047-0.070}$ \\
\hline $100\Omega_bh^2$ &
$2.33^{+ 0.10+0.20+0.32}_{- 0.09-0.17-0.25}$ (*) \\
\hline
A & $0.837$  \\
\hline
$n_s$ & $0.978^{+ 0.028+0.058+0.084}_{- 0.022-0.041-0.059}$ \\
\hline
$H_0$ & $70.7^{+ 2.4+4.9+7.4}_{- 2.3-4.6-6.6}$ \\
\hline
$\tau$ & $0.152^{+ 0.067+0.127+0.146}_{- 0.056-0.101-0.136}$ \\
\hline
$\omega_0 $ & $-0.981^{+ 0.193+0.38+0.57}_{- 0.193-0.37-0.52}$ \\
\hline
$\omega_1 $ & $-0.05^{+0.65+1.13+1.38}_{-0.83-1.92-2.88}$ \\
\hline
\end{tabular}
\end{center}
* $\Omega_b = 0.0466$.

The value we obtained for $1/c$ with uncertainties given
respectively by  $68.26\%$, $95.44\%$ and $99.73\%$, is \bear
\frac{1}{c} &=& 0.56^{+0.12+0.28+0.41}_{-0.20-0.48-0.55}\nonumber
\eear

For equation of state $II$ we used  the parameters given in table
2.

\textbf{Table 2}: cosmological parameters for equation of state
$II$
\\
\begin{center}
\begin{tabular}{|c|c|}
\hline
$\Omega_{\Lambda }$ & $0.76^{+ 0.03+0.6}_{- 0.13-0.20}$ \\
\hline
$\Omega_bh^2$ & $0.0266^{+ 0.0004}_{- 0.0066}$ (*) \\
\hline
A & $1.22$ \\
\hline
$n_s$ & $1.1^{+ 0.0}_{- 0.2}$ \\
\hline
$H_0$ & $73^{+ 5}_{- 12}$ \\
\hline
$\tau$ (reionization depth) & $0.35^{+ 0.04}_{- 0.33}$ \\
\hline
$\omega_0 $ & $-1.48^{+0.78+1.05}_{-0.19-0.45}$ \\
\hline
$\omega_1 $ & $3.86^{+ 0.29+1.01}_{- 4.6-6.36}$ \\
\hline
\end{tabular}
\end{center}
*$\Omega_b = 0.05$.\\

The value we obtained for $1/c$ with uncertainty given by
$68.26\%$, $95.44\%$ and $99.73\%$, respectively, is \bear
\frac{1}{c} &=& 0.42^{+0.06+0.16+0.27}_{-0.24-0.39-0.42}\nonumber
\eear

\begin{figure}[t]
   \centering
     \subfigure[\hspace{0.1cm}]
     {\includegraphics[width=0.4\textwidth]{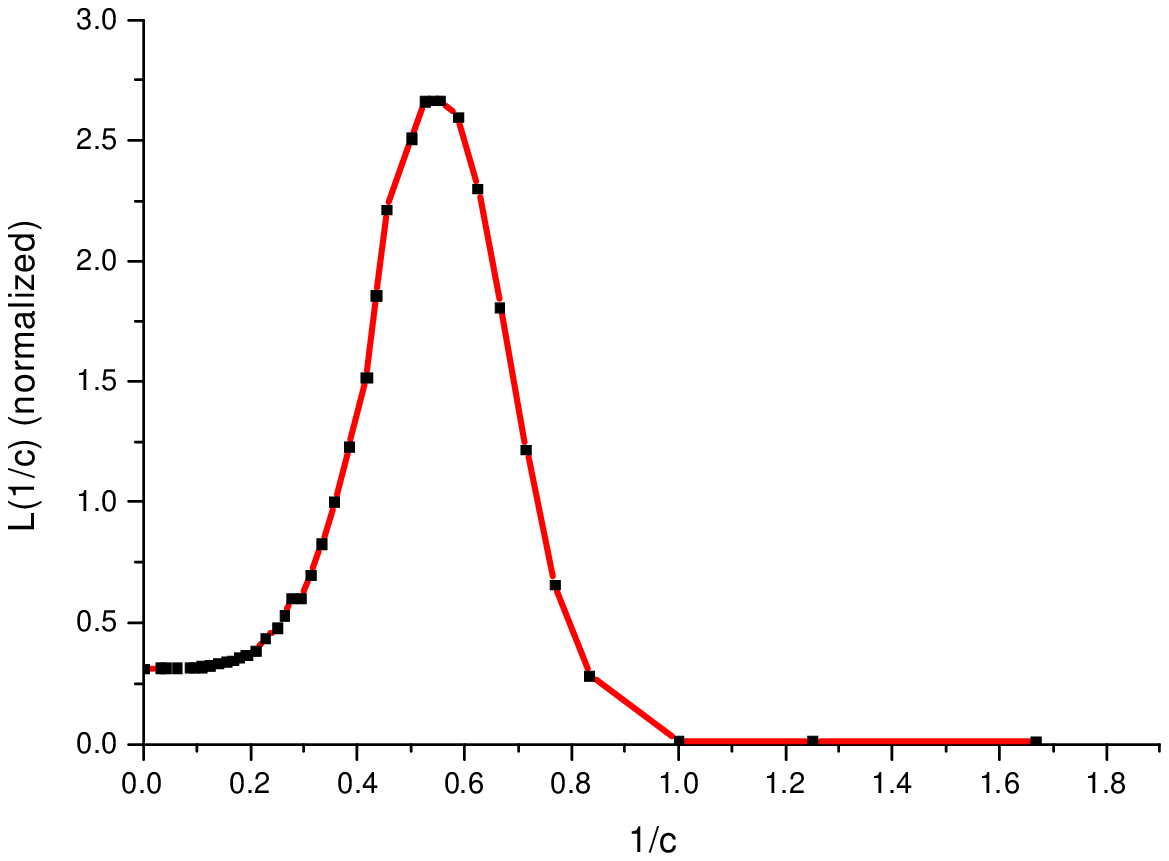}}
     \hspace{0.01\textwidth}
     \subfigure[\hspace{0.1cm}]
     {\includegraphics[width=0.4\textwidth]{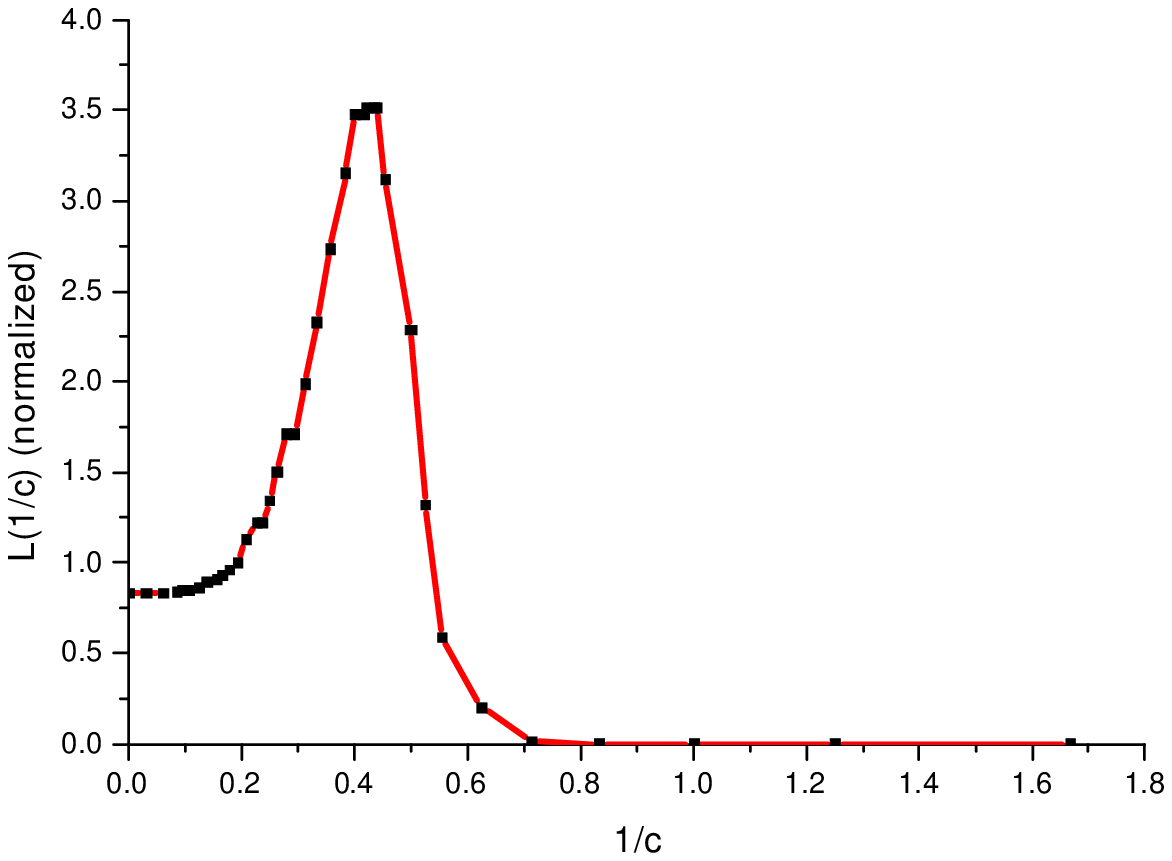}}
     \caption{Likelihood functions for $1/c$.
     The first figure corresponds to equation of state I and
     the second to equation of state II. Note that in the case
     of the equation of state II $c$ has to be larger, $c_{min}\sim 1.4$.}
\label{likelihoodc}
\end{figure}

The two sets of values for $1/c$ above are already compatible at
level $1\ \sigma$. Since the parameter $c$ affects only the region
of small multipoles the calculation above could be performed with
some simplifying assumptions concerning the other cosmological
parameters, which are defined by the region with high multipoles.
The corresponding likelihood functions are given in figure
(\ref{likelihoodc}) ($1/c$ was used because for numerical
precision and normalization).

After the determination of $c$ for the two equations of state for
dark energy, as described above, we estimated the coupling between
dark energy and dark matter --- $b^2$ by comparing
(\ref{eqestint}) with the parametrizations (\ref{eqest1}) and
(\ref{eqest2}). We have to find the maximum of the likelihood
function $\mathcal{L}(b^2,1/c,\Omega_{\Lambda}, \Omega_b, H_0,
n_s, \frac{dn_s}{dlnk}, \tau, \omega_0, \omega_1, A)$ in order to
find the best value of $b^2$ and its error bars \cite{verde}.

For the first parametrization model,  we have \bear
  - \frac{1}{3} - \frac{2\sqrt{\Omega_{\Lambda}(z)}}{3c} -
\frac{b^2}{\Omega_{\Lambda}(z)} &\simeq& \omega_0^{I} +
\omega_1^{I}\frac{z}{1+z}  \quad . \eear For the second
parametrization, we get \bear
  - \frac{1}{3} - \frac{2\sqrt{\Omega_{\Lambda}(z)}}{3c} -
\frac{b^2}{\Omega_{\Lambda}(z)} &\simeq& \omega_0^{II} +
\omega_1^{II}\frac{z}{(1+z)^2}  \quad . \eear

\begin{figure}[t]
   \centering
     \subfigure[\hspace{0.1cm}]
     {\includegraphics[width=0.4\textwidth]{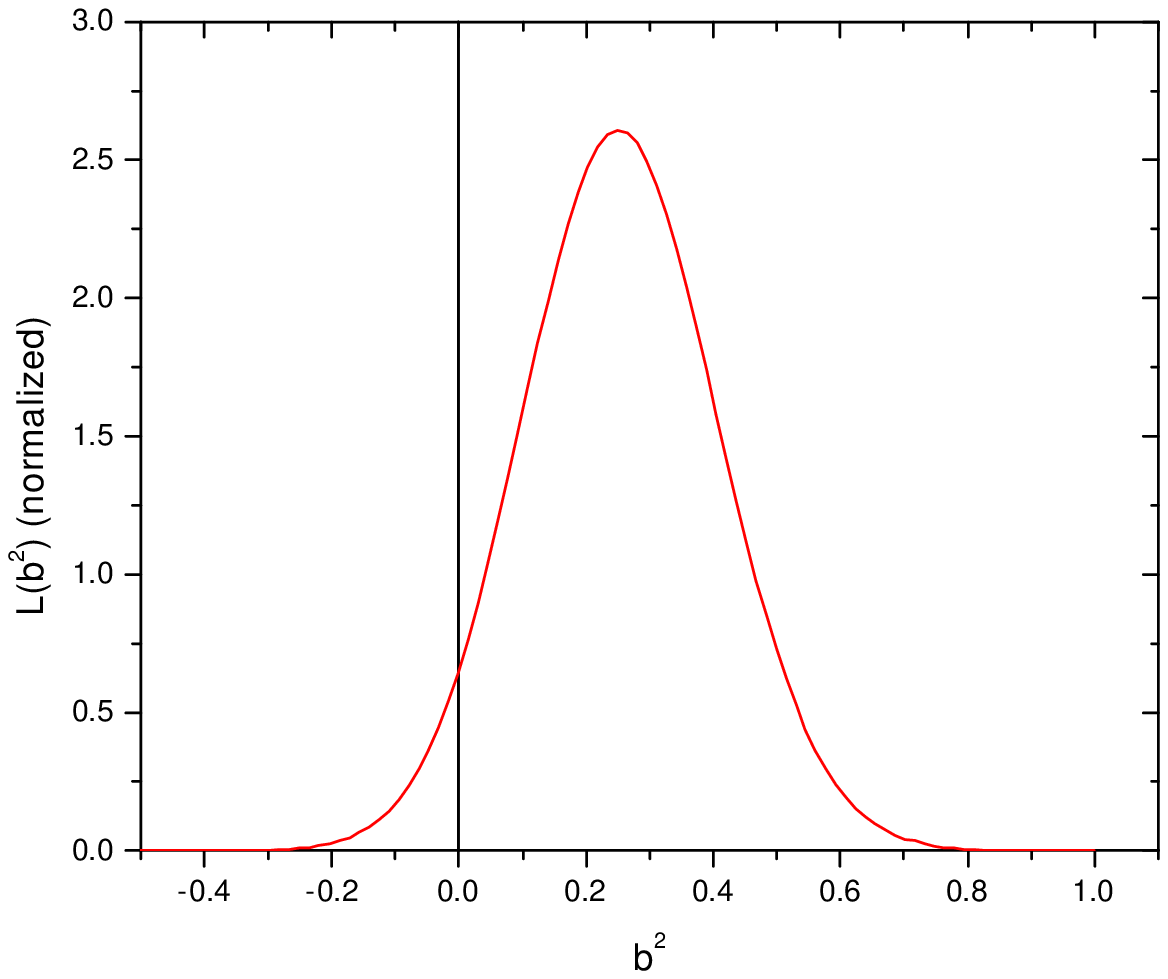}}
     \hspace{0.01\textwidth}
     \subfigure[\hspace{0.1cm}]
     {\includegraphics[width=0.4\textwidth]{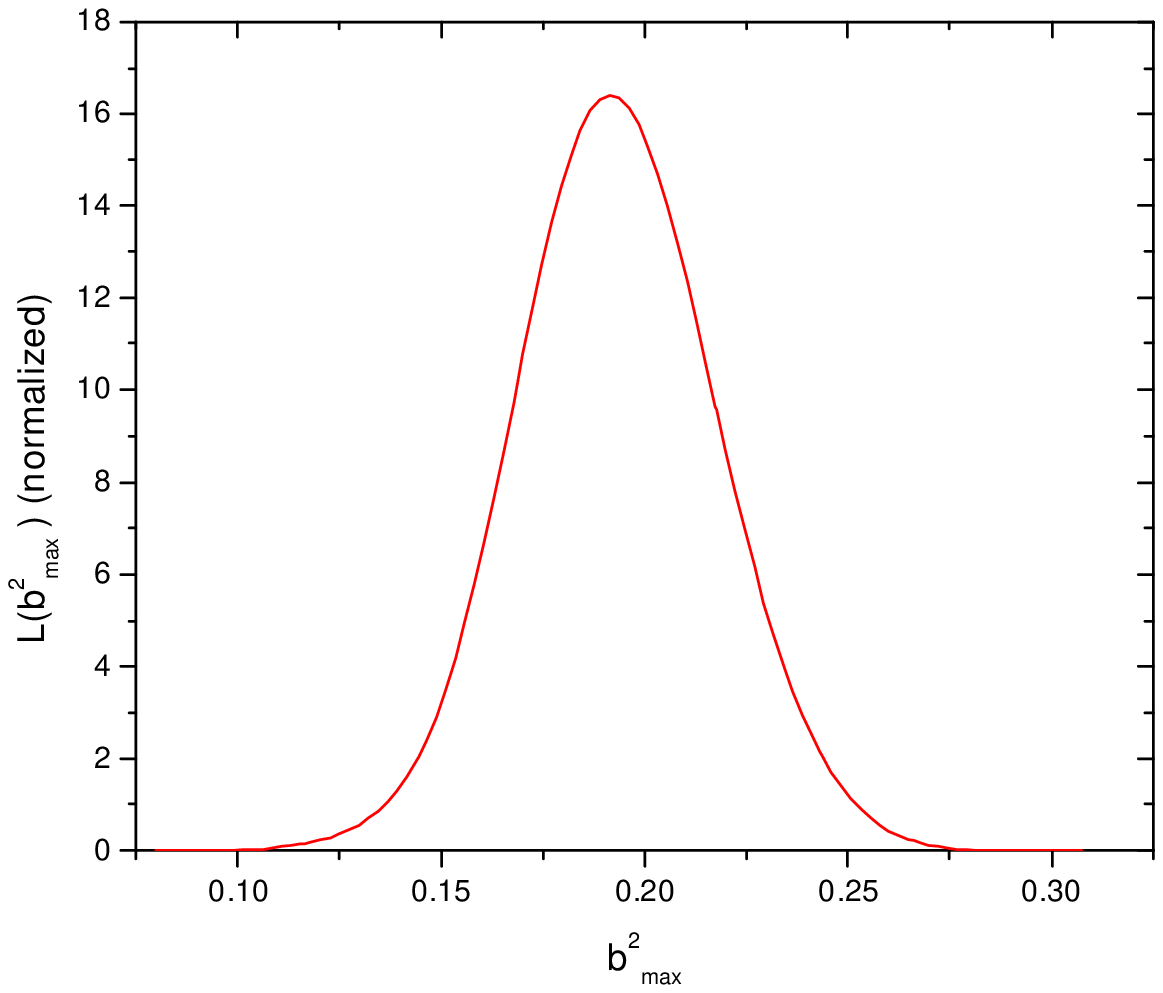}}
     \caption{Likelihood functions for $b^2$ and $b^2_{max}$.
     These figures correspond to equation of state I.}
\label{likelihoodb}
\end{figure}

Today ($z=0$) the above two equations boil down to \bear
\label{omega0}
  - \frac{1}{3} - \frac{2\sqrt{\Omega_{\Lambda}(0)}}{3c} -
\frac{b^2}{\Omega_{\Lambda}(0)} &\simeq& \omega_0    \quad . \eear
Isolating $b^2$ we have \bear \label{b2} b^2 &\simeq&
-\Omega_{\Lambda}(0){\Bigg(} \omega_0 + \frac{1}{3} +
\frac{2\sqrt{\Omega_{\Lambda}(0)}}{3c} {\Bigg)} \quad . \eear We
can now evaluate $b^2$ and the respective uncertainty. The
likelihood function for $b^2$ as well as for $b^2_{max}$ defined
below appear in figure (\ref{likelihoodb}) for parametrization I.

The results obtained for $b^2$ using the two parametrizations with
uncertainties  $68.26\%$, $95.44\%$ and $99.73\%$ are given below,
\bear b^2 &=& 0.25^{+0.16+0.32+0.48}_{-0.15-0.30-0.45} \ (eq.\ of\
state\ 1)\quad ,\nonumber \eear and \bear b^2 &=&
0.7^{+0.1+0.5+0.8}_{-0.6-0.9-1.3}\ (eq.\ of\ state\ 2)\nonumber
\quad . \eear

  We can impose a further constraint in $b^2$, using the condition
  given in equation (8) of \cite{wga}, that is, that
  $\Omega_D^\prime$ is always positive, namely that the
  dark energy has
  an increasing role with the flow of time,
\be b^2 < b^2_{max}  \frac{1-\Omega_{\Lambda}}{3}
(1+2\frac{\sqrt{\Omega_{\Lambda}}}{c})  \quad . \ee

  The values of $b^2_{max}$ using parametrization 1, with the
  respective intervals of uncertainty $68.26\%$, $95.44\%$ and
  $99.73\%$ are
\bear b^2_{max} &=&
0.191^{+0.027+0.052+0.075}_{-0.022-0.047-0.075} \quad .\nonumber
\eear

Combining $b^2$ and $b^2_{max}$, \bear
0.10 < b^2 < 0.22\quad (1\ \sigma),\nonumber\\
-0.05 < b^2 < 0.24\quad (2\ \sigma),\nonumber\\
-0.20 < b^2 < 0.27\quad (3\ \sigma).\nonumber \eear

For parametrization 2, we obtain \bear b^2_{max} &=&
0.14^{+0.06+0.10+0.15}_{-0.03-0.08-0.12}\quad .\nonumber \eear
Combining $b^2$ and $b^2_{max}$ for parametrization 2 we get \bear
0.1 < b^2 < 0.2\quad    (1\ \sigma)\quad,\nonumber\\
-0.2 < b^2 < 0.24\quad   (2\ \sigma)\quad ,\nonumber\\
-0.6 < b^2 < 0.29\quad   (3\ \sigma)\quad .\nonumber \eear

As a conclusion we see that the parameter $b^2$ is non zero with a
confidence of 90\%. Using the prior that the dark energy always
increases, we have, for the probability the $0.05 < b^2 < 0.2$ is
of the order of 75\%.

We computed the probability of finding a non-positive value of
$b^2$ according to the above data and found it to be of the order
4.5\%.

Finally, let us comment on the improvement of the model when
comparing it to the $\Lambda$CDM model. We have first to call
attention to the fact that the first modification, namely the
holographic input, leads us out of the  $\Lambda$CDM case,
requiring modifications in the equations thus used. Nevertheless,
comparing the low $\ell$ part of the spectrum only, the $\chi^2$
comes down to roughly one half (the full $\chi^2$ shows little
difference due to the broadness of the problem). The values for
the spectrum corresponding to the first eight (8) multipoles are
\bear
\chi^2_{\Lambda CDM}&=&8.3\quad ,\\
\chi^2_{par\; 1}&=&5.2\quad ,\\
\chi^2_{par\; 2}&=&2.8\quad . \eear Such values imply a nice
improvement with respect to the $\Lambda$CDM model, completing our
picture which tried, and in our view succeeded, in arguing that
interaction between Dark Energy and Dark Matter are essential on
the top of natural in a Quantum Field Theory description.

\section{Conclusions and overview}
We discussed  several observational constraints on the two
parameter space of holographic dark energy/dark matter model. We
can confidently conclude that the interaction must be nonvanishing
in order that we explain all available data at the same time.

Concerning the holography coefficient $c$, we know that it must
have a lower limit due to the second law of thermodynamics as well
as from the low $\ell$ CMB data. We actually have considered the
holographic model in all possible scenarios, including the
$\Lambda CDM$ and the likelihood function always leads to a $c$
larger than unit.

The age constraint provided by the old quasar leads to a lower
limit of the $b$ parameter, namely $b^2>0.05$, which seems to be
an outcome which should be respected as foreseen by all
observations we analyzed. For the likelihood function discussed in
section 5, we found that $b^2$ is larger than 0.05 with
probability 0.9 and with the prior $b<b_{max}$ with probability of
the order 0.75. The age limit also puts maximum values for the
same parameter. For the age today we got a maximum $b$ of the
order $b^2<0.2$, which is also the range obtained in section 5 in
case we take into account the requirement of an always increasing
dark energy, or $b^2 \lesssim b_{max}^2\approx 0.2$.

Therefore all observational constraints point to the same range
$0.05 < b^2 < 0.2$.

For the case of the $c$ parameter the bounds are not so tight. The
CMB data show very confidently that $c>1$ in order that not too
high values of $\ell$ are suppressed in the power spectrum, an
independent check of the result already obtained from more formal
arguments. However, the likelihood function is very spread from
unity to infinity, and we are not in position to say much more. We
thus worked most of the time with the limiting value $c=1$.

Although extremely simple, the model opens up wide possibilities
towards new physics: 30\% of the energy is composed of a new state
of matter, interacting with the rest 67\% yet unknown, a
fascinating and probably extremely complex universe!

Certainly new kinds of interaction have to be considered,
especially in case they can be computed from first principles and
more well defined methods of theoretical physics and quantum field
theory. Work in this direction is under way \cite{holo,strings}.
We thus conclude pointing out that, in our opinion, interaction
shoud be an essential worry in the description of the dark sector
of the Universe.

Acknowledgement: The work of E.A. and S.M. is partially supported
by CNPq (Conselho Nacional de Desenvolvimento Cientifico e
Tecnologico) and FAPESP (Fundacao de Ampara a Pesquisa do Estado
de Sao Paulo). The work of B. Wang was partially supported by NNSF
of China, Ministry of Education of China and Shanghai Education
Commission. The work of C.-Y. L. was supported in part by the
National Science Council under Grant No. NSC- 93-2112-M-259-011.

\end{document}